\documentclass[12pt,onecolumn, draftclsnofoot]{IEEEtran}
\usepackage[cmex10]{amsmath}
\usepackage{cite}
\usepackage{algorithm}
\usepackage{algorithmic}
\usepackage{array}
\usepackage{mdwtab}
\usepackage[tight,footnotesize]{subfigure}
\usepackage[dvips]{graphicx}
\usepackage{subfigure}
\usepackage{amsfonts}

\begin{document}
%\graphicspath{{eps/}}
\DeclareGraphicsExtensions{.eps}
\title{Compressed Sensing SAR Imaging with Multilook Processing}
\author{Jian Fang, Zongben Xu, Bingchen Zhang, Wen Hong, Yirong Wu,
% <-this % stops a space
\thanks{J. Fang and Z. B. Xu are with the School of Mathematics and Statistics, Xi'an Jiaotong University, Xi'an 710049, China (Email of corresponding author: zbxu@mail.xjtu.edu.cn).}% <-this % stops a space
\thanks{B. C. Zhang, W. Hong, Y. R. Wu are with the Institute of Electronics, Chinese Academy of Sciences, Beijing 100190, China.}% <-this % stops a space

% <-this % stops a space and the National Natural Science Foundations of China (Grants No. 61075054, 60975036, 11171272)
%\thanks{none}% <-this % stops a space
}

\maketitle

\begin{abstract}
Multilook processing is a widely used speckle reduction approach in synthetic aperture radar (SAR) imaging. Conventionally, it is achieved by incoherently summing of some independent low-resolution images formulated from overlapping subbands of the SAR signal. However, in the context of compressive sensing (CS) SAR imaging, where the samples are collected at sub-Nyquist rate, the data spectrum is highly aliased that hinders the direct application of the existing multilook techniques. In this letter, we propose a new CS-SAR imaging method that can realize multilook processing simultaneously during image reconstruction. The main idea is to replace the SAR observation matrix by the inverse of multilook procedures, which is then combined with random sampling matrix to yield a multilook CS-SAR observation model. Then a joint sparse regularization model, considering pixel dependency of subimages, is derived to form multilook images. The suggested SAR imaging method can not only reconstruct sparse scene efficiently below Nyquist rate, but is also able to achieve a comparable reduction of speckles during reconstruction. Simulation results are finally provided to demonstrate the effectiveness of the proposed method.

\end{abstract}


% \fi
\IEEEpeerreviewmaketitle

\section{Introduction}
\IEEEPARstart{S}{ynthetic} aperture radar (SAR) is a high-resolution active microwave radar imaging system that is widely used in both military and civilian applications\cite{Cumming2004}. However, the state-of-art SAR imaging systems with increasing resolution and swath require more and more amount of measurements that imposes great burden on the storage and downlink bandwidth.

The recent development of compressed sensing (CS)\cite{Candes2006cs}\cite{Donoho2006} brings the possibility of reconstructing sparse scene with far fewer measurements than that Nyquist requires. Several applications appear in the last few years. For instance, in \cite{Herman2009}, the high-resolution radar was designed by emitting incoherent signals with reduced size. In \cite{Patel2010}, a general CS-SAR model was proposed by discretizing the SAR observation function exactly into an observation matrix, and solving by CS algorithms. The CS was also applied to SAR tomography\cite{Zhu2010} to achieve super resolution along elevation direction. In \cite{Ender2010}\cite{Potter2010}, a variety of possible applications were summarized to show the great potential of CS in future development of radar imaging techniques. All these works strongly demonstrated the exclusive advantages of CS-SAR on relaxing the required measurements in improving the image quality \cite{Cetin2001}.

On the other hand, multilook processing\cite{Cumming2004}\cite{porcello1976} is a widely used speckle reduction method in conventional SAR signal processing, which is achieved by dividing the signal spectrum and then incoherently averaging the recovered subimages. It is distinguished from some other advanced speckle reduction methods, like Lee filter\cite{lee1991}, wavelet method\cite{xie2002}, TV regularization \cite{shi2008}, as an in-process method, in spite of the drawback of degraded resolution.

In this letter, we consider the realization of multilooking in CS-SAR imaging. The purpose is to enrich the function of CS-SAR algorithms, and meanwhile trying to use in-process speckle reduction to avoid the darkened targets (seriously caused by speckles) being inappropriately shrinked even truncated during CS reconstruction. To achieve the goal, a nature consideration is to integrate multilooking into the construction of CS-SAR observation model. However, the difficulty addresses in the fact that the current CS-SAR approaches depend on a time-domain model that is not well compatible with frequency-domain operations. To resolve this issue, we substitute the SAR observation matrix in CS-SAR model by the inverse of multilooking procedures, where the time-frequencies transformations build a bridge between multilooking and compressed sampling while at the same time bring a faster computation. Then, by discovering the sparsity of the summed multilook data, a block sparse regularization problem\cite{Daubechies2004,Yuan2006}, considering the joint support of all the looks, is established to recover multilook subimages from the new model. The derived images are finally averaged to achieve the speckle reduction. Accordingly, we can obtain an efficient CS-SAR imaging method that integrates the advantages of both CS and multilook processing.

\section{Background Knowledge of CS-SAR and Multilook Processing}
%In this section, we introduce the CS-SAR observation model with multilook expression. Firstly, the preliminary knowledge of CS-SAR, multilook processing, as well as the challenging of their combination are summarized. An approximated observation model is then constructed to connect them using the inverse of the look formation procedure.

\subsection{Compressed Sensing SAR Imaging}
The main characteristic of CS-SAR, as compared with traditional SAR, is its capacity of relaxing the required measurements for reconstruction, by introducing a random subsampling\cite{Patel2010} process during data acquisition. This problem is generally described using the following linear problem:
\begin{equation}\label{eq:ob2d}
{\bf y}_s={\bf \Theta}{\bf y}={\bf \Theta} \textbf{H} {\bf x}+{\bf n}
\end{equation}
where $\bf y$ is the fully sampled raw data, ${\bf y}_s$ is the compressed measurements, ${\bf \Theta}$ denotes the compressed sampling matrix, $\bf x$ is the scene to be recovered, $\bf n$ is the additional noise, $\bf H$ is the SAR observation matrix. Denoting the azimuth and range time of the $m$th sample in $\bf y$ as $\eta_m,\tau_m$, the $n$th resolution cell in ${\bf x}$ as $\eta_{n},\tau_{n}$, we have ${\bf H}_{i,j}=h({\eta}_{n}-{\eta}_{m},{\tau}_{n}-{\tau}_{m})$(more details can be seen in \cite{yang2013}), where

\begin{eqnarray}\label{eq:ck}
h(\eta,\tau)&=&\omega_a(\eta)\exp [ - j4\pi f_0\frac{R(\eta)}{c}]\\
&&\omega_r(\tau-\frac{2R(\eta)}{c})\exp[j\pi K_r(\tau-\frac{2R(\eta)}{c})^2]
\end{eqnarray}
In (\ref{eq:ck}), $\omega_a,\omega_r$ are respectively the azimuth and range envelope function, $c$ is the velocity of light, $f_0$ is the carrier frequency, $R$ is the slant range, $K_r$ is the FM rate of signal.

When the scene is sparse, say, most of the entries of $\bf x$ are zero, and the sensing matrix ${\bf A}={\bf \Theta} \textbf{H}$ satisfies specific conditions like RIP\cite{Candes2006}, the theory of CS guarantees that $\bf x$ can be exactly recovered from $\bf y_s$ with the following $L_q$ ($0<q\leq 1$) optimization:
\begin{equation}\label{eq:CSq}
\min\limits_{\bf x}{\kern 5pt}\|{\bf x}\|_q {\kern 5pt}s.t.{\kern 5pt}{\bf y}_s = \textbf{A} {\bf x}
\end{equation}
where $\|{\bf x}\|_q=(\sum_i|x_i|^q)^{1/q}$ is the $L_q$ quasi-norm. By solving (\ref{eq:CSq}), the SAR image can be recovered with much few measurements than Nyquist rate requires, thereby has great potentials to reduce the complexity of the SAR system.

\subsection{Multilook Processing}
As is known that, speckle in SAR images\cite{Oliver2004} arises from coherent sum of scatters within a resolution cell and is manifested as randomly multiplicative "noise" on magnitudes. To suppress them, multilook processing is an effective approach that involves an incoherent summing of some independent subimages. More specifically, multilook processing consists of two major procedures, which are respectively the look formation step and look summation step. In look formation step, it generates a couple of low-resolution complex images using different subbands of the data. Because of the linear dependance between azimuth time and Dopplers, each subband corresponds to a different beam looking angle (but is always difficult to extracted directly in time-domain), hence the corresponding part of data are called "looks"\footnote{To make the suggested method concise, we assume multilook processing on azimuth only, while the extension on range is quite similar.}. In the summation step, all looks are averaged by summing the magnitude square, and then perform a square root\cite{Cumming2004}
\begin{equation}\label{eq:sum}
{\bf z}(k)=\sqrt{\sum_{i=1}^{L}{|{\bf x}_i(k)}|^2}
\end{equation}
where $L$ denotes the number of looks. From a statistical view, the multilook process maintains the mean but reduce the variance of the target intensity, so tat can suppress the speckle effect.

In this letter, we aim at the combination of CS-SAR imaging and multilooking. The motivations can be explained in the following three folds. Firstly, the multilooking is the most simple and practical speckle reduction method that is worthy to be embed in CS-SAR imaging.  Secondly, the CS algorithms always penalizes much more on small targets. As a result, the targets darkened by speckles are liable to be misidentified and truncated during reconstruction. For this reason, it is worthy to provide a in-process speckle reduction method so as to alleviate such problem. Finally, we have held some preliminary experiments, for example the Fig. \ref{fig_simud}, which shows that the speckles may be deteriorated as the sampling rate decreases. The reason may be that the CS-SAR observation model (\ref{eq:ob2d}) with finite grid can not completely fit the fully speckle case, and the loss of measurements will decrease the effective number of looks. With these considerations, we attempt to introduce a multilook CS-SAR imaging method in the next section.

\section{Compressed Sensing SAR Imaging with Multilook Processing}
\subsection{Multilook CS-SAR Observation Model}
\begin{figure}[t]

\centering
\subfigure[\ ]{\includegraphics[width=0.8\textwidth]{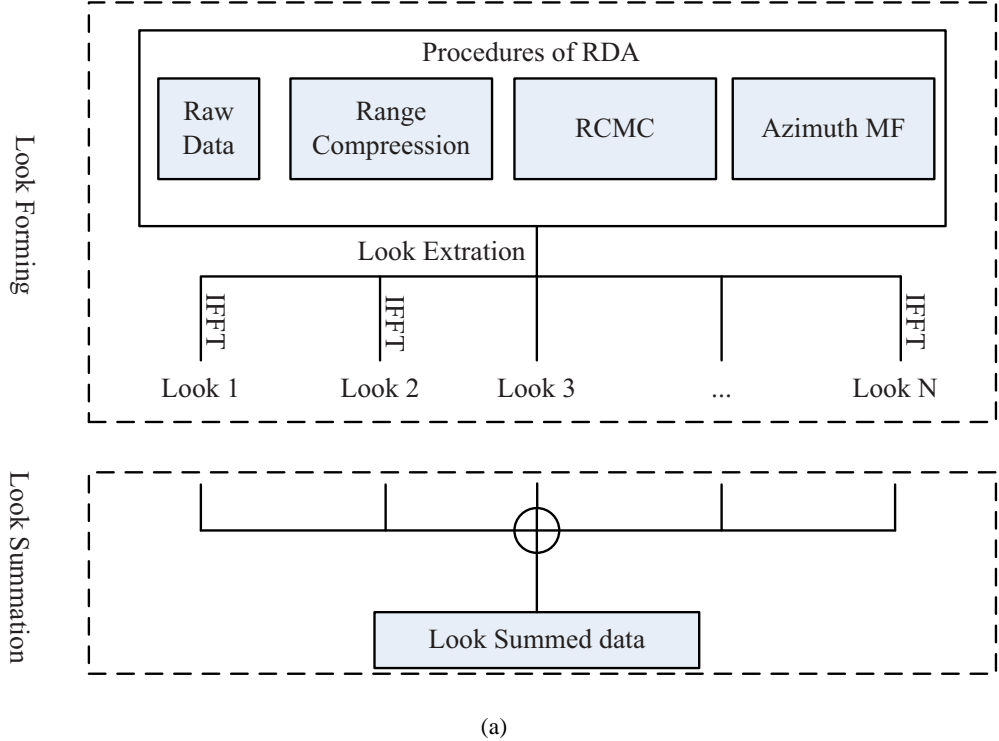}}
%\label{figfirstcase}}
\hfil
\caption{The main procedures of multilook processing based on RDA.}
\label{fig_ml}
%\label{fig_sim}
\end{figure}

Considering the look formation step, which is denoted by $\rm M$, the subimages are focused by
\begin{equation}
{\bf X}=[{{\bf x}_1,...,{\bf x}_L}]={\rm M}({\bf y})
\end{equation}
where $\bf X$ stacks all subimages ${\bf x}_i$ into a matrix, ${\rm M}$ is always constructed from high precision methods, for example the range Doppler algorithm (RDA) and chirp-scaling algorithm, that can be computed very efficiently.

To obtain a CS observation model like (\ref{eq:ob2d}), one certain way is to look for the possibilities that $\rm M$ is invertible, so that $\bf y$, further the compressed measurements ${\bf y}_s$, can be expressed via multiple looks
\begin{equation}\label{eq:mlo}
{\bf y}_s=\Theta {\rm M}^{-1}({\bf X})+{\bf n}
\end{equation}
where ${\rm M}^{-1}$ (or the general inverse in many cases), if exists, is the multilook observation that we suggest to apply in CS multilook imaging. Fortunately, we found that ${\rm M}^{-1}$ is obtainable under some weak assumptions, i.e., the subbands are nonoverlapping. This constraint is introduced to avoid indeterminacy in the shared parts of looks such that $\bf y$ can be uniquely determined by $\bf X$. Then, we will take range-Doppler algorithm (RDA) as an example to show how the inverse can be constructed.

As shown in Fig. \ref{fig_ml}, the look forming procedure based on RDA includes generally four steps, which are consequently the range compression, range cell migration cracking, azimuth matched filtering and look extraction. The first three steps follow the standard RDA while in look extraction step, the spectrum is divided and inverse IFFT is applied on each look to form the subimages. In general, the procedures $\rm M$ formally has the following expression:
\begin{equation}\label{eq:apoh}
{\bf x}_i={\bf F}_{a'}^{\rm H}{\rm R}_i({\bf Q}{\bf D}{\bf F}_{a}{\bf P}{\bf y})
\end{equation}
where ${{\bf F}_{a}}$ is the discrete Fourier transformation matrix on azimuth, ${{\bf F}_{a'}}$ is the inverse Fourier transformation with length $\frac{1}{L}$ of ${\bf{F}}_{a}$, $\rm D$, ${\rm R}_i,{\bf P},{\bf Q}$ are matrices denoting respectively the range cell migration crack (realized in sinc-polation), the nonoverlapping look extraction, range matched filter and azimuth matched filter.

To derive the inverse of the $\rm M$, we can take the inverse of every sub-procedures. The inverse procedures of range compression (in frequency form), RCMC and azimuth matched filtering can be approximated according to \cite{Fang2012} by:
\begin{equation}\label{eq:IRDA}
{\bf y}={\bf P}^{\rm H}{\bf F}^{\rm H}_{a}{\bf C}{\bf Q}^{\rm H}{\bf s}
\end{equation}
where $\bf s$ is the image spectrum, $\bf C$ is matrix for range cell migration, which is approximated obtained by reversing the RCMC through interpolation. It is worthy noting that as will be seen in Fig. \ref{fig_algo}, all these operations are in practice realized efficiently via FFTs, phase multiplication and interpolation, and the matrix form is convenient for method formulation.

%\footnote{In practice, (\ref{eq:IRDA}) is realized straightforwardly using matched filters and FFTs, some of the details can be found in \cite{Fang2013}}.

Then, according to the assumption that the subbands are disjoint, the complete image spectrum can be represented by taking FFTs on each look and stacking them in order:
\begin{equation}\label{eq:stack}
{\bf s} = {\rm S}({\bf X})={\rm vec} \left( \begin{array}{l}
{{\bf{F}}_{a'} }{{\bf{X}}_1}\\
{{\bf{F}}_{a'} }{{\bf{X}}_2}\\
 \cdot  \cdot  \cdot \\
{{\bf{F}}_{a'} }{{\bf{X}}_L}
\end{array} \right)
\end{equation}
where ${\bf X}_i$ (${\bf x}_i={\rm vec}({\bf X}_i)$) is the two-dimensional image of the $i$th look. Further, with (\ref{eq:IRDA}), (\ref{eq:stack}) and (\ref{eq:ob2d}), the compressed SAR data could be expressed by
\begin{equation}\label{eq:apo}
{\bf y}_s=\Theta{\rm G}({\bf X})+n=\Theta {\bf P}^{\rm H}{\bf F}^{\rm H}_{a}{\bf C}{\bf Q}^{\rm H}{\rm S}({\bf X})+n
\end{equation}
where we use $\rm G$ to denote the inverse of the look forming step. It is easily verified that $\rm G$ is a linear operator, and the transition conjugate of $\rm G$ is just $\rm M$ (see \cite{Fang2012}), which brings much convenience in constructing the algorithms in the next section.

{\bf Remarks} Firstly, to better understand the multilook CS (MLCS) model, we can decompose (\ref{eq:IRDA}) into three parts. The $\bf \Theta$ and $\bf S$ at each end are the sampling matrix and the spectrum slitting procedure, which are the key procedures of CS and ML respectively. The operators ${\bf P},{\bf C}$ and $\bf Q$ come from the approximated observation proposed in \cite{Fang2012} and have the following two functions. Firstly, all these operators are implemented in frequency domain that can be realized with $\mathcal{O}(n\log n)$ computation, much faster than time-domain correlation (nearly $\mathcal{O}(n^2)$). Second and more importantly, the operators acts as a transition between time-domain operator $\bf \Theta$ and frequency domain operator $\bf S$, therefore combines smoothly the CS and ML methods to form a new observation.

Secondly, by observing that ${\bf X}={\rm M}({\bf y})$, we actually target on the reconstruction of the matched filter result in undersampling case by using the MLCS model. In this sense, the speckle reduction can always be achieved whenever a successful reconstruction can be reached.

Finally, the method is not limited to the form of RDA, but can be generalized similarly to many other well-developed reconstruction algorithms, like Chirp-Scaling Algorithm (CSA), $\omega-{\rm k}$ algorithm.

\subsection{The Reconstruction Method}
For a proper CS-SAR application, the target scene is required to be sparse or compressible. In single-look case, it is usually assumed that most entries of the complex data array $\bf x$ defined in (\ref{eq:ob2d}) are zero or negligible. In multilook case, however, we are interested in the summed image $\bf z$, and the sparsity can be defined by
\begin{equation}\label{eq:gsra}
\|{\bf z}\|_1=\|{\bf X}\|_{2,1}=\sum\limits_{j=1}^n \| {\bf X}^{j}\|_2
\end{equation}
%It should be noted that the such sparsity is more strict because $\bf z$ is not always sparse when all ${\bf x}_i$ are sparse.
where $\|\cdot\|_{2,1}$ is a mixed norm. The form coincides with the definition of block sparsity, which promotes sparsity along rows and requires the variables in the active row should be seen together. This definition is quite rationale since wherever there exists targets in a resolution cell, it can always be observed in all looks at the same position.
\begin{figure*}[htbp]
\centering
\includegraphics[width=0.9\textwidth]{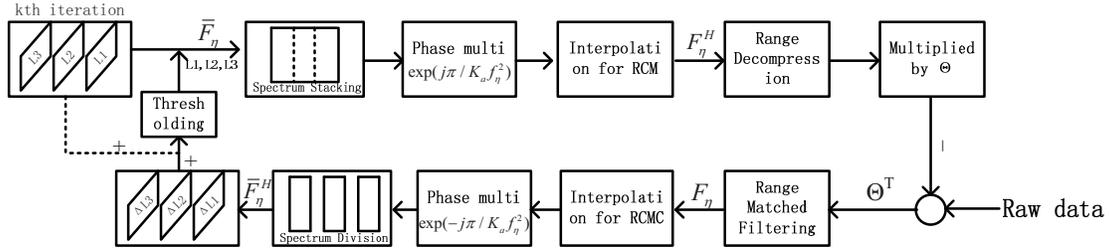}
%\label{figfirstcase}}
\caption{The diagram for implementation of the MLCS algorithm based on the use of RDA.}
\label{fig_algo}
%\label{fig_sim}
\end{figure*}

With (\ref{eq:apo}) and (\ref{eq:gsra}), we can acquire the following multilook CS-SAR reconstruction model:
\begin{equation}\label{eq:model}
\min_\textbf{X} \{ \|\textbf{y}_s-\Theta{\rm G}({\bf X})\|_F^2+\lambda\|{\bf X}\|_{2,1}\}
\end{equation}
To solve (\ref{eq:model}), there exists very efficient algorithm based on iterative thresholding algorithm\cite{Bengio2009}, which takes the following iterative scheme:
\begin{equation}\label{eq:gth}
{{\bf{X}}^{(k + 1)}} = {{\rm{E}}_{2|1,\lambda \mu }}\left( {{{\bf{X}}^{(k)}} + \mu {\Theta ^{\rm{T}}}{\rm{J}}\left( {{{\bf{y}}_s} - \Theta {\rm{G}}({\bf{X}})} \right)} \right)
\end{equation}
where the group thresholding operator ${\rm{E}}_{2|1,\tau}$ ($\tau=\lambda\mu$) operate on each row of ${\bf X}$ independently, say, ${\rm E}({\bf X})=[{\rm e}({\bf X}^{1})^{\rm T},...,{\rm e}({\bf X}^{n})^{\rm T}]^{\rm T}$ and
\begin{equation}\label{eq:gthresholder}
\rm{e}_{2|1,\tau} ({\bf x})=\frac{e_{1,\tau}(\|{ {\bf x}}\|_2)}{\|{{\bf x}}\|_2}{ {\bf x}}
\end{equation}
in (\ref{eq:gthresholder}) $e_{1,\tau}(x) ={\rm sgn}(x) \max(|x|-\tau,0)$ is the so-called soft-thresholder. The iteration can be understood as alternatively a matching step that extract useful information from the residue, and the thresholding step that suppresses alias and enables sparsity. After obtaining the subimages ${\bf x}_i$, they are summed to form the final speckled reduced image.

For convenience of use, we draw the diagram of the implementation as in Fig. \ref{fig_algo}.

%\begin{algorithm}\label{eq:th2d}
%\caption{: Iterative thresholding algorithm based multilook CS-SAR imaging}
%\label{alg:ISTA}
%\begin{algorithmic}[1]
%{\small
%\REQUIRE
%   Compressed SAR raw data $\textbf{y}_s$,
%\ENSURE
%   The summed multilook image $\textbf{z}^*$
%\renewcommand{\algorithmicensure}{\textbf{Initial:}}
%\ENSURE
%    $\textbf{X}^{(0)}=0$,$\lambda,\mu$ and max iteration $I_{\max}$
%\FOR{$i=0$ to $I_{\max}$}
%\STATE  Residue: ${\textbf{r}}^{(i)}=\textbf{y}_s-{\bf \Theta} {\rm G}(\textbf{X}^{(i)})$
%\STATE  MF on residue:$\Delta \textbf{X}^{(i)}={\rm J}({\bf \Theta}^T \textbf{r}^{(i)})$
%\STATE Thresholding: $\textbf{X}^{(i+1)}={\rm E}_{2|1,\lambda\mu}(\textbf{X}^{(i)}+\mu\Delta \textbf{X}^{(i)})$
%\ENDFOR
%\STATE Look Summation: ${\bf z}(i)=\| {\bf X}^{i}\|_2$
%}
%\end{algorithmic}
%\end{algorithm}

From the previous subsections, we can see that the suggested method has constituted an efficient multilook CS-SAR imaging method. While preserving CS features, the new method is capable of reducing speckle noise. Moreover, the imaging procedures are cost-saving that is possible to be applied in high dimensional SAR applications. We will provide simulations in the next section to further support these benefits.

\section{Simulation Results}
In this section, several simulations are provided to evaluate the CS reconstruction ability as well as the speckle reduction capacity of the proposed SAR imaging method. They are measured respectively by the whether successful sub-Nyquist reconstruction can be achieved and the equivalent look number (ENL) of the obtained image, defined as $ENL=\left(\frac{\mu_L}{\sigma_L}\right)^2$\cite{Cumming2004}, where $\mu_L$ is the mean and $\sigma_L$ is the standard deviation of the intensity of the interested region.

\subsection{Experiment Setup}
In all simulations, the scene was taken as $150\times 150$, while in the region of $24\times 24$ at the center of the scene, we independently draw 400 Rayleigh distributed points targets randomly located within each resolution cell (totally 2.3 million targets). The SAR parameters were taken as the slant range $R=20 {\rm km}$, the radar velocity $v=350 {\rm m/s}$, the radar center frequency $f_c=5 {\rm GHz}$, range FM rate $B_r=75{\rm MHz}$, the pulse duration $T_r=2\mu {\rm s}$ and zero beam squint angle. The data were firstly generated by exact slant range with Gaussian noise of 20 dB and were then randomly sampled with different rate to yield compressed measurements. For the reconstruction method, the regularization parameter $\lambda$ and maximum iteration were set respectively with $0.02L$ and 500.

\subsection{Simulation Results}
\begin{figure}[t]
\centering
\subfigure[\ ]{\includegraphics[clip=true,trim=30 30 30 30,width=0.4\textwidth]{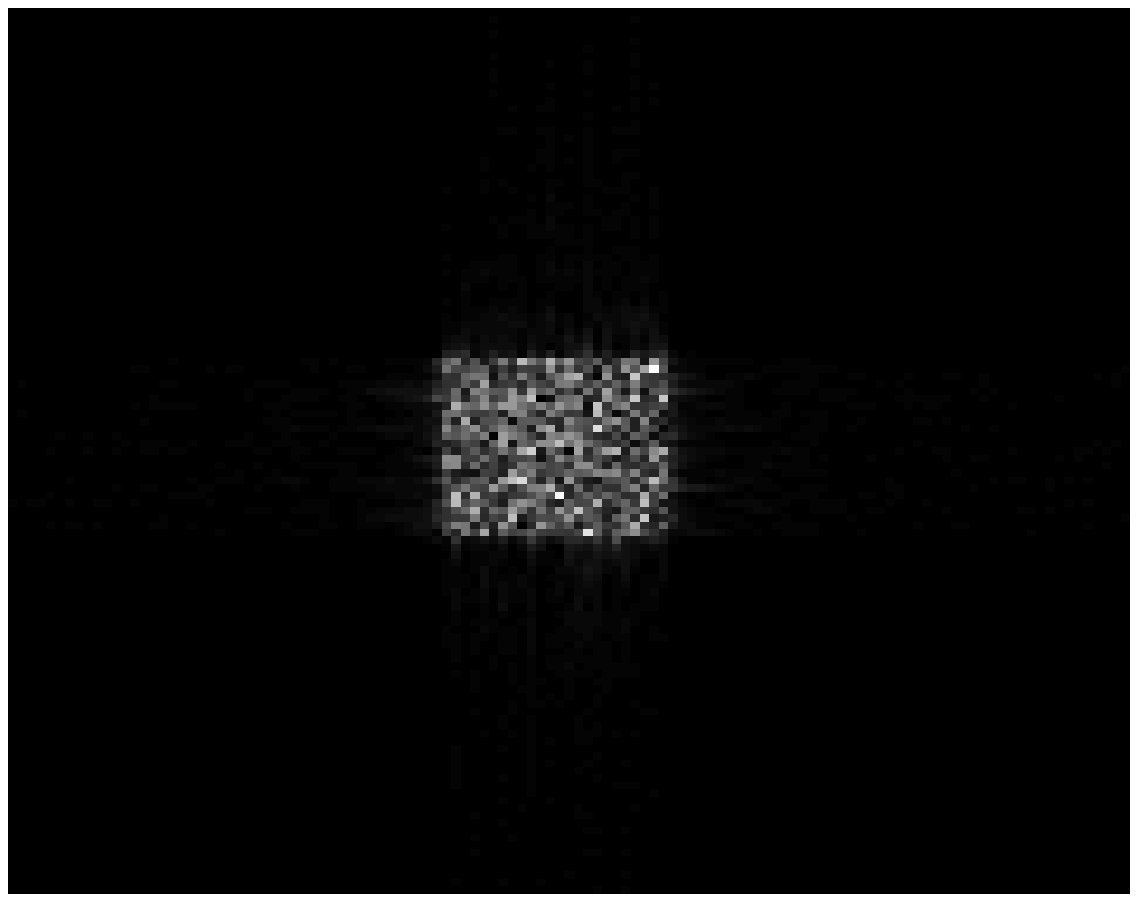}}
%\label{figfirstcase}}
\hfil
%\label{fig_second_case}}}
\subfigure[\ ]{\includegraphics[clip=true,trim=30 30 30 30,width=0.4\textwidth]{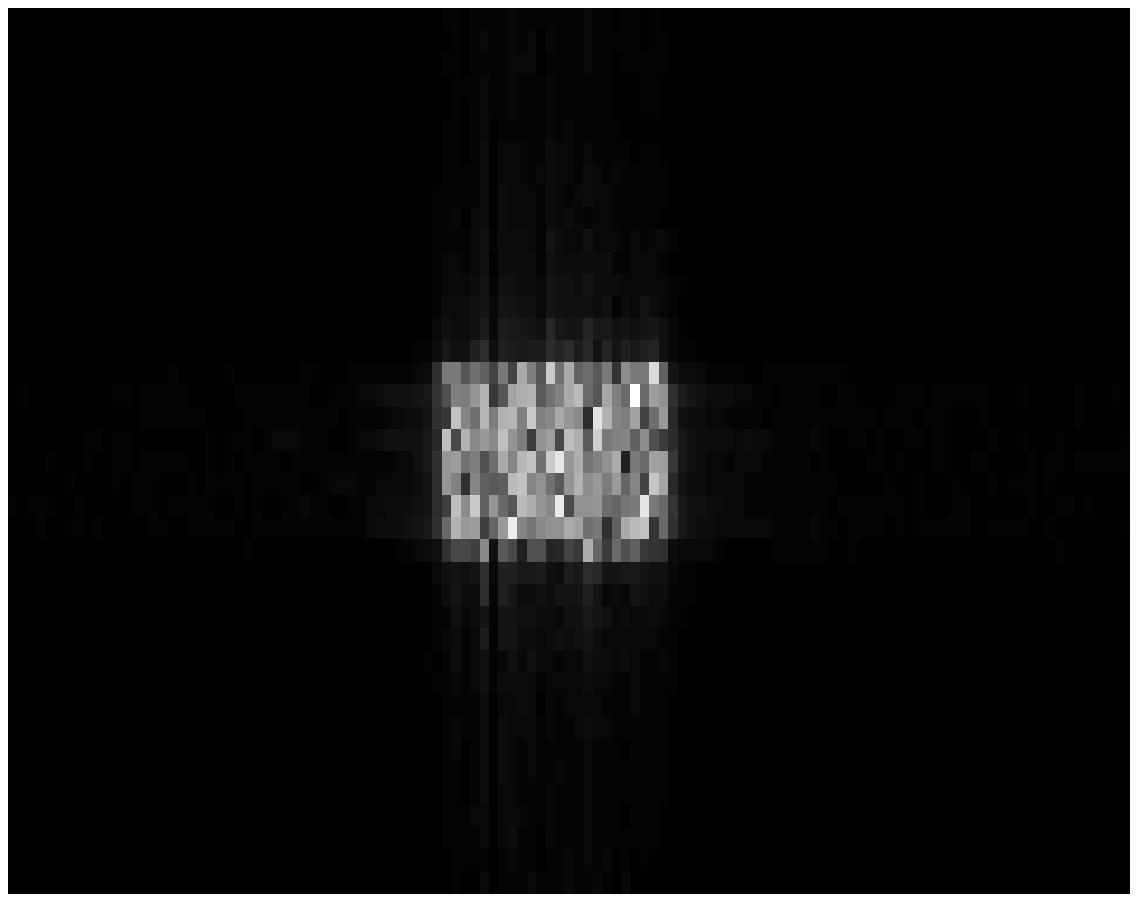}}
\hfil
\subfigure[\ ]{\includegraphics[clip=true,trim=30 30 30 30,width=0.4\textwidth]{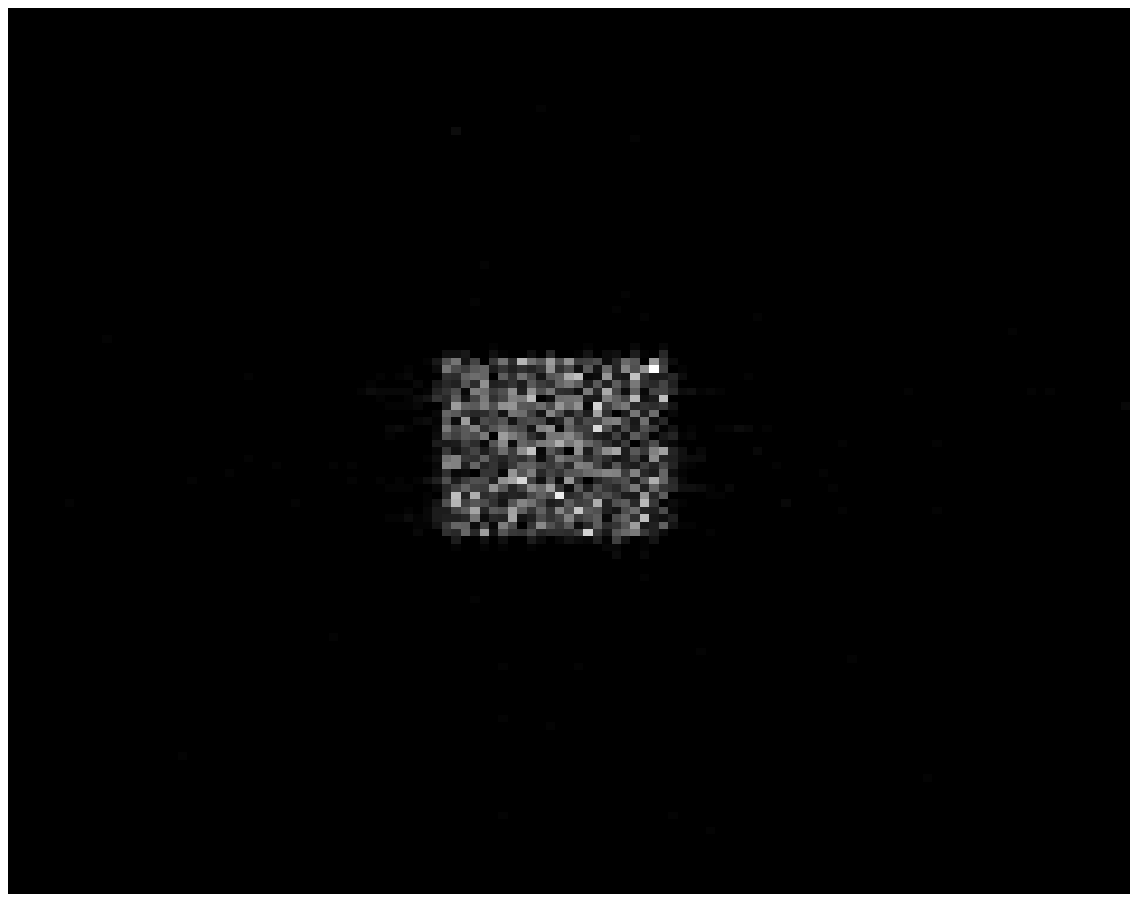}}
\hfil
\subfigure[\ ]{\includegraphics[clip=true,trim=30 30 30 30,width=0.4\textwidth]{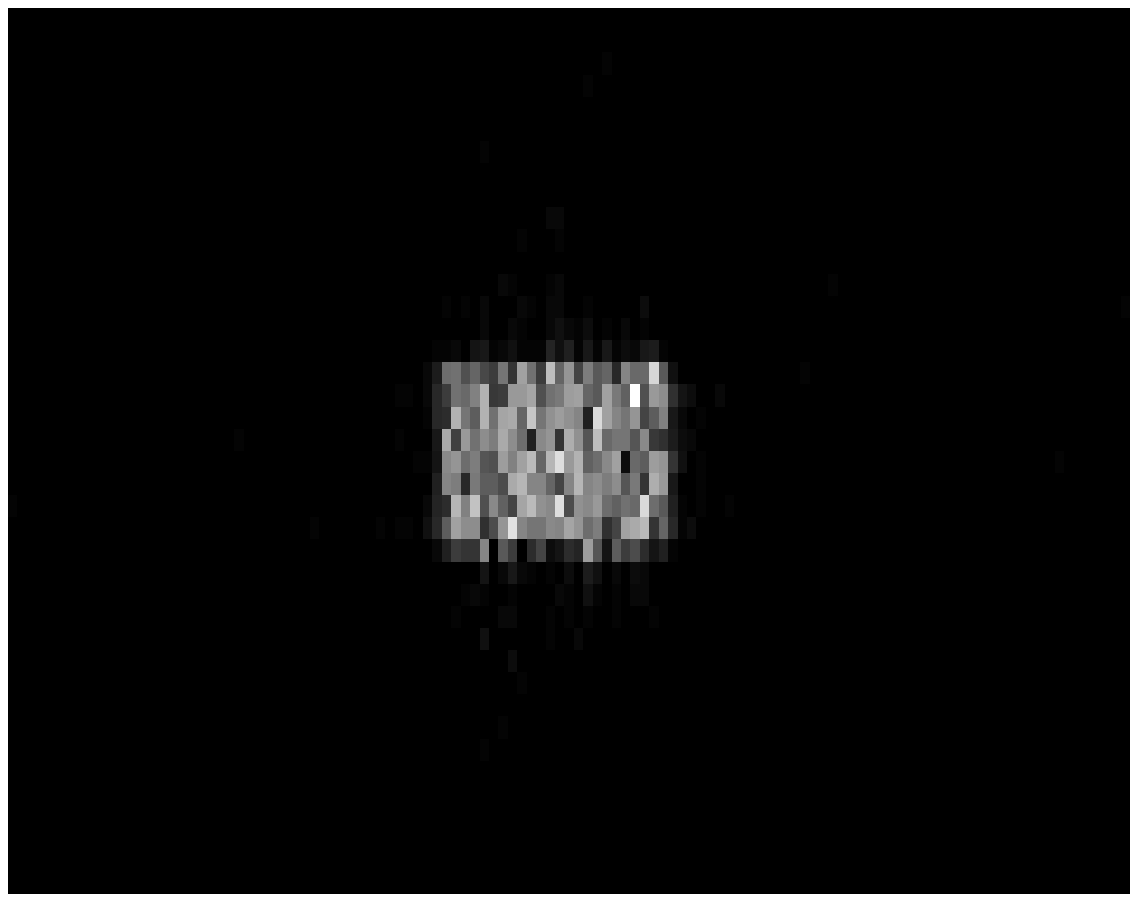}}
\hfil
\caption{The multilook reconstruction results of a $24\times24$ region. From left to right corresponds to 1 and 3 looks. And the top row presents RDA results with full samples while the bottom shows the reconstruction of the proposed method with 20\% samples.}
\label{fig_simut}
%\label{fig_sim}
\end{figure}

\begin{figure}[t]
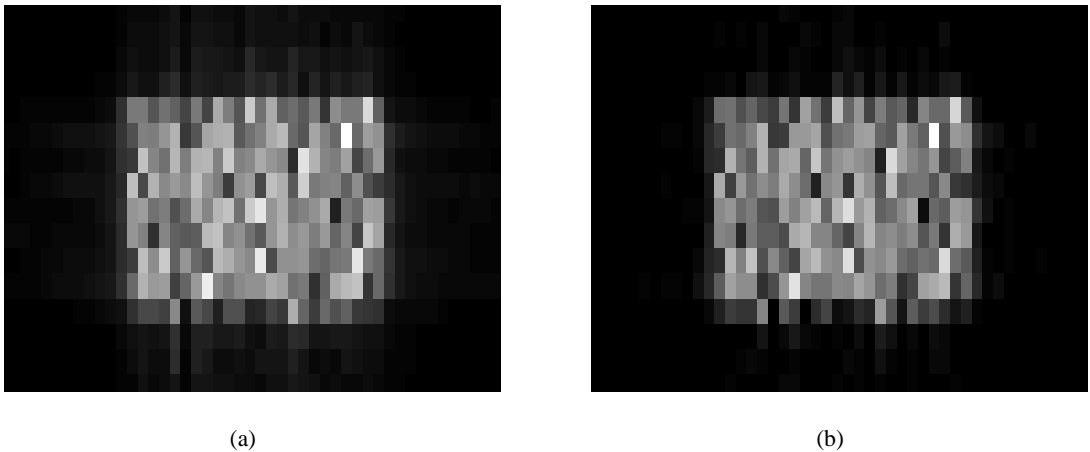

\centering
\subfigure[\ ]{\includegraphics[clip=true,trim=95 80 105 80,width=0.4\textwidth]{R3}}
%\label{figfirstcase}}
\hfil
%\label{fig_second_case}}}
\subfigure[\ ]{\includegraphics[clip=true,trim=95 80 105 80,width=0.4\textwidth]{C3}}
\caption{Detail comparison on RDA and MLCS with 3 looks.}
\label{fig_simutd}
%\label{fig_sim}
\end{figure}

\begin{figure}[t]
\centering
\subfigure[\ ]{\includegraphics[width=0.4\textwidth]{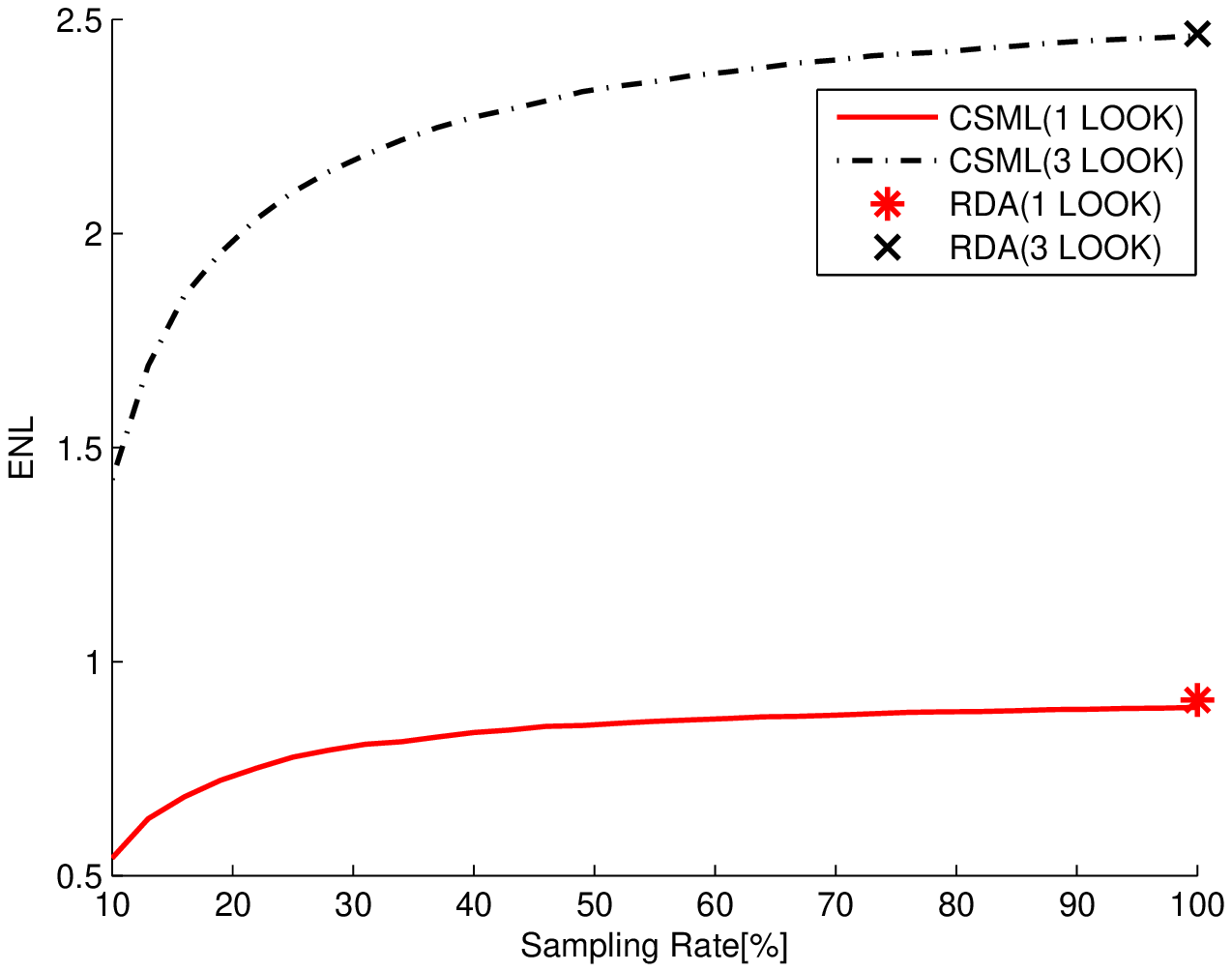}}
%\label{figfirstcase}}
\hfil
\subfigure[\ ]{\includegraphics[width=0.4\textwidth]{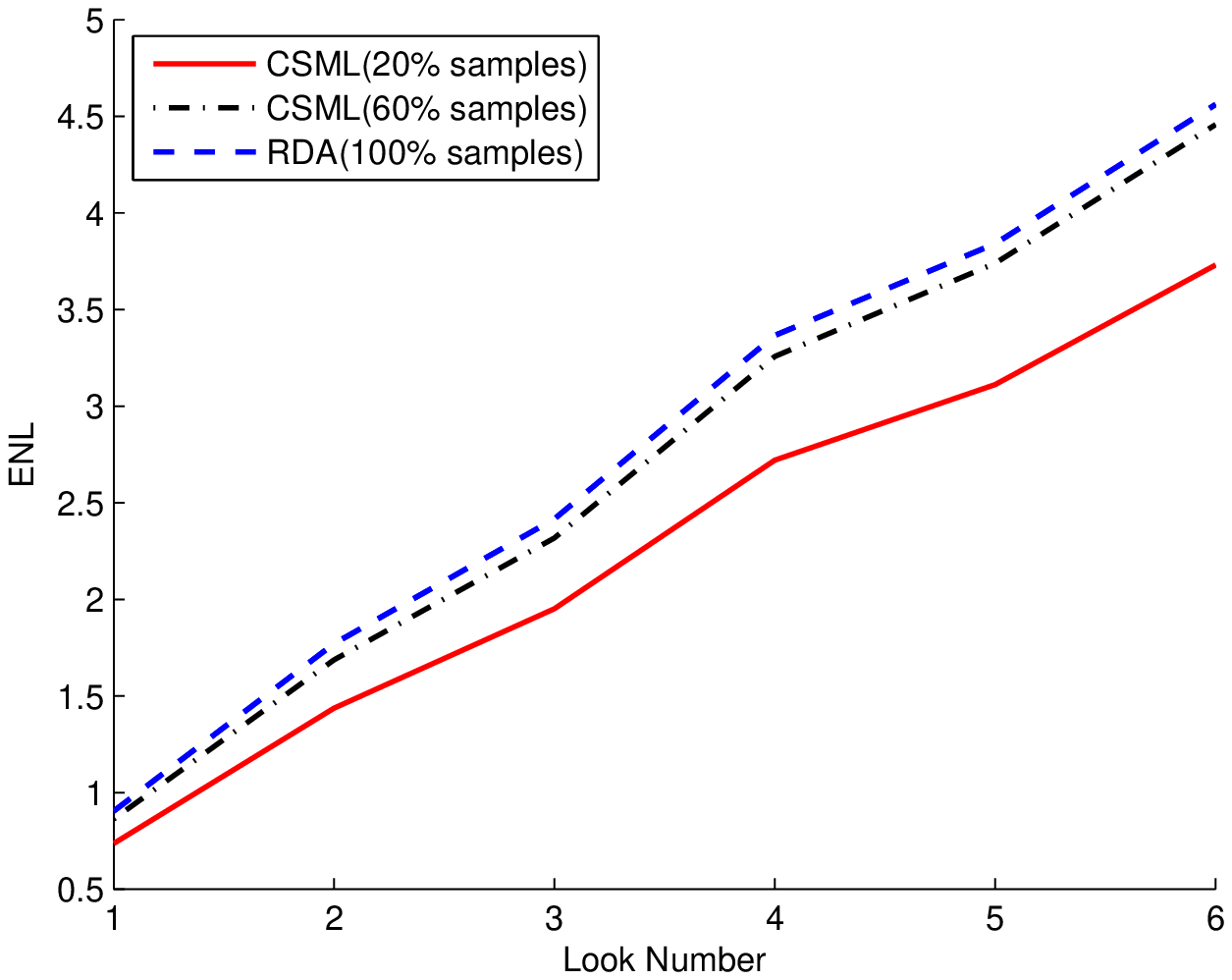}}
\hfil
\caption{The ENL as a function of sampling rate (left) and look number (right).}
\label{fig_simud}
%\label{fig_sim}
\end{figure}

In Fig. \ref{fig_simut}, we compare the CS reconstruction results of the proposed multilook CS-SAR imaging method with 20\% samples and the multilook RDA outputs using full data. It is seen that the proposed method successfully recovered the interested region in undersampling case without artifacts. Meanwhile, for both methods, the speckle effect is reduced as the number of look increases. More details comparison in Fig. \ref{fig_simutd} shows that the CS multilook outputs appear almost as same as that of the matched filter result but the proposed method, benefited form sparse regularization, was shown with reduced sidelobe.

In Fig. \ref{fig_simud}, we show the relationship between ENL, sampling rate and look number for the proposed method. More specifically, Fig. \ref{fig_simud}(a) draws the mean ENL calculated from 100 independent experiments as a function of the sampling rate. It is seen that with full samples, the RDA and the proposed method achieves a similar ENL. But as the sampling rate decreases, the ENL slightly declines. Then in Fig. \ref{fig_simud}(b), we can see that as the look number increases, the ENL of both RDA and the proposed method promotes. Moreover, the proposed method with higher sampling rate provides better performance.

All of the above results show that the proposed method can reconstruct sparse scene with far less samples than Nyquist rate requires. Meanwhile, the speckles can be effectively reduced during the reconstruction progress, but there is a little loss on ENL when the sampling rate decreases.

\section{Conclusion}
In this paper, we have proposed a novel CS-SAR imaging method with multilooking function, with which speckle reduction can be realized simultaneously with sub-Nyquist reconstruction.

The main contributions of the present work are as follows. First, we proposed a multilook CS observation model by taking the inverse of the look forming step to embed multilook processing in SAR. Second, a joint sparse regularization model is presented to solve the problem.

It is worthwhile, however, to remark that although speckle reduction in CS-SAR can be brought, it is found in simulations that the reconstruction with CS-SAR may deteriorate ENL, especially when sampling rate is very low. Thus, how and what extent do the sampling way and rate affect the reconstruction deserves a further study.

%and the National Natural Science Foundations of China
%(Grants No. 61075054, 60975036, 11171272)
\ifCLASSOPTIONcaptionsoff
  \newpage
\fi



%\bibliographystyle{IEEEtran}

\bibliographystyle{./IEEEtran}
\bibliography{./IEEEabrv,./paper}

%\begin{biography}[{\includegraphics[width=1in,height=1.25in,clip,keepaspectratio]{mshell}}]{Michael Shell}

%\end{IEEEbiography}


%
%% insert where needed to balance the two columns on the last page with
%%\newpage


%\vfill


\end{document}